\documentclass[prd,twocolumn,reprint,preprintnumbers,nofootinbib]{revtex4-1}

\input{header}

\begin{document}
\preprint{OSU-HEP-20-14}
\title{Neutral current neutrino interactions at \texorpdfstring{\fasernu}{fasernu}}

\author{Ahmed Ismail}
\email{aismail3@okstate.edu}
\affiliation{Department of Physics, Oklahoma State University, Stillwater, Oklahoma 74078}

\author{Felix Kling}
\email{felixk@slac.stanford.edu}
\affiliation{Theory Group, SLAC National Accelerator Laboratory, Menlo Park, California 94025}

\author{Roshan Mammen Abraham}
\email{rmammen@okstate.edu}
\affiliation{Department of Physics, Oklahoma State University, Stillwater, Oklahoma 74078}

\begin{abstract}
In detecting neutrinos from the Large Hadron Collider, FASER$\nu$ will record the most energetic laboratory neutrinos ever studied. While charged current neutrino scattering events can be cleanly identified by an energetic lepton exiting the interaction vertex, neutral current interactions are more difficult to detect. We explore the potential of FASER$\nu$ to observe neutrino neutral current scattering $\nu N \to \nu N$, demonstrating techniques to discriminate neutrino scattering events from neutral hadron backgrounds as well as to estimate the incoming neutrino energy given the deep inelastic scattering final state. We find that deep neural networks trained on kinematic observables allow for the measurement of the neutral current scattering cross section over neutrino energies from 100 GeV to several TeV. Such a measurement can be interpreted as a probe of neutrino non-standard interactions that is complementary to limits from other tests such as oscillations and coherent neutrino-nucleus scattering.
\end{abstract}

\maketitle

\section{Introduction}
\label{sec:intro}

As the only neutral and uncolored fermions in the Standard Model (SM), neutrinos are perhaps some of the least well understood particles in nature. Precision measurements of neutrino interactions across a variety of energy scales are important in order to understand neutrino oscillations as well as to probe new physics in the neutrino sector. However, most experiments studying neutrinos from artificial sources are limited to maximum energies of a few hundred GeV. The exception is the Large Hadron Collider (LHC): As the highest energy particle collider ever built, the LHC is the source of the most energetic neutrinos created in a controlled laboratory environment. Proton-proton collisions typically lead to a large number of hadrons produced in the far-forward direction, which can inherit a significant fraction of the protons' momenta. The decays of these hadrons then lead to an intense and strongly collimated beam of high-energy neutrinos of all three flavors along the beam collision axis. While the possibility of probing neutrinos at the LHC was discussed as early as 1984~\cite{DeRujula:1984pg, Vannucci:253670, DeRujula:1992sn, Park:2011gh,  Buontempo:2018gta, Feng:2017uoz}, no LHC neutrino has been detected yet. This situation will change soon with the upcoming FASER$\nu$ detector~\cite{Abreu:2019yak, Abreu:2020ddv}, which is expected to detect thousands of neutrino interactions during Run 3 of the LHC. 

One of the most basic measurements involving neutrinos is the scattering cross section of neutrinos off matter. Both charged current (CC) and neutral current (NC) scattering offer sensitive tests of the SM (for a review, see Ref.~\cite{Formaggio:2013kya}).  The majority of neutrino cross section measurements are from experiments using terrestrial sources at low energies, extending up to neutrino energies of about $300~\gev$~\cite{Kodama:2007aa, Tanabashi:2018oca}. Astrophysical neutrino cross sections have also been measured at IceCube~\cite{Aartsen:2017kpd, Bustamante:2017xuy}, probing very high neutrino energies between $\sim 10~\tev$ and $\sim 1$~PeV albeit with significant uncertainties. At the few $100~\gev$ to a few TeV scale, there are no precise measurements of neutrino scattering. FASER$\nu$ offers an opportunity to study neutrinos at these energies. The ability of FASER$\nu$ to measure CC neutrino scattering has been studied, but it is not known yet how well NC scattering could be measured with LHC neutrinos. In this work, we fill this gap by studying the capability of FASER$\nu$ to determine the neutrino NC cross section at the LHC.

NC scattering is significantly more difficult to identify than its CC counterpart. While CC scattering produces an outgoing lepton that carries much of the original neutrino energy, neutrino NC interactions result only in a neutrino and any products of the recoiling nucleus. In FASER$\nu$, there are significant backgrounds to NC scattering from neutral hadron interactions within the detector. We simulate neutrino scattering and neutral hadron events at FASER$\nu$, and use a neural network to effectively separate signal from background using kinematic information of the final state. By applying another neural network, we show that the energy of the neutrino can be estimated with $\sim 50\%$ uncertainty. Taken together, we find that FASER$\nu$ will be able to measure the neutrino NC cross section as a function of neutrino energy.

The rest of this paper is organized as follows. In Section~\ref{sec:faserv}, we provide an overview of FASER$\nu$ and how it probes neutrino scattering.  Section~\ref{sec:simulation} describes our simulation of signal, background and the detector. Then, we detail our analysis procedure in Section~\ref{sec:analysis}. Section~\ref{sec:results} contains the results of this analysis, including our estimate of the precision with which FASER$\nu$ could measure the NC scattering cross section and our interpretation in terms of limits on neutrino non-standard interactions (NSI). We conclude in Section~\ref{sec:outlook}.

\begin{figure*}[th!]
\includegraphics[width=1.0\textwidth]{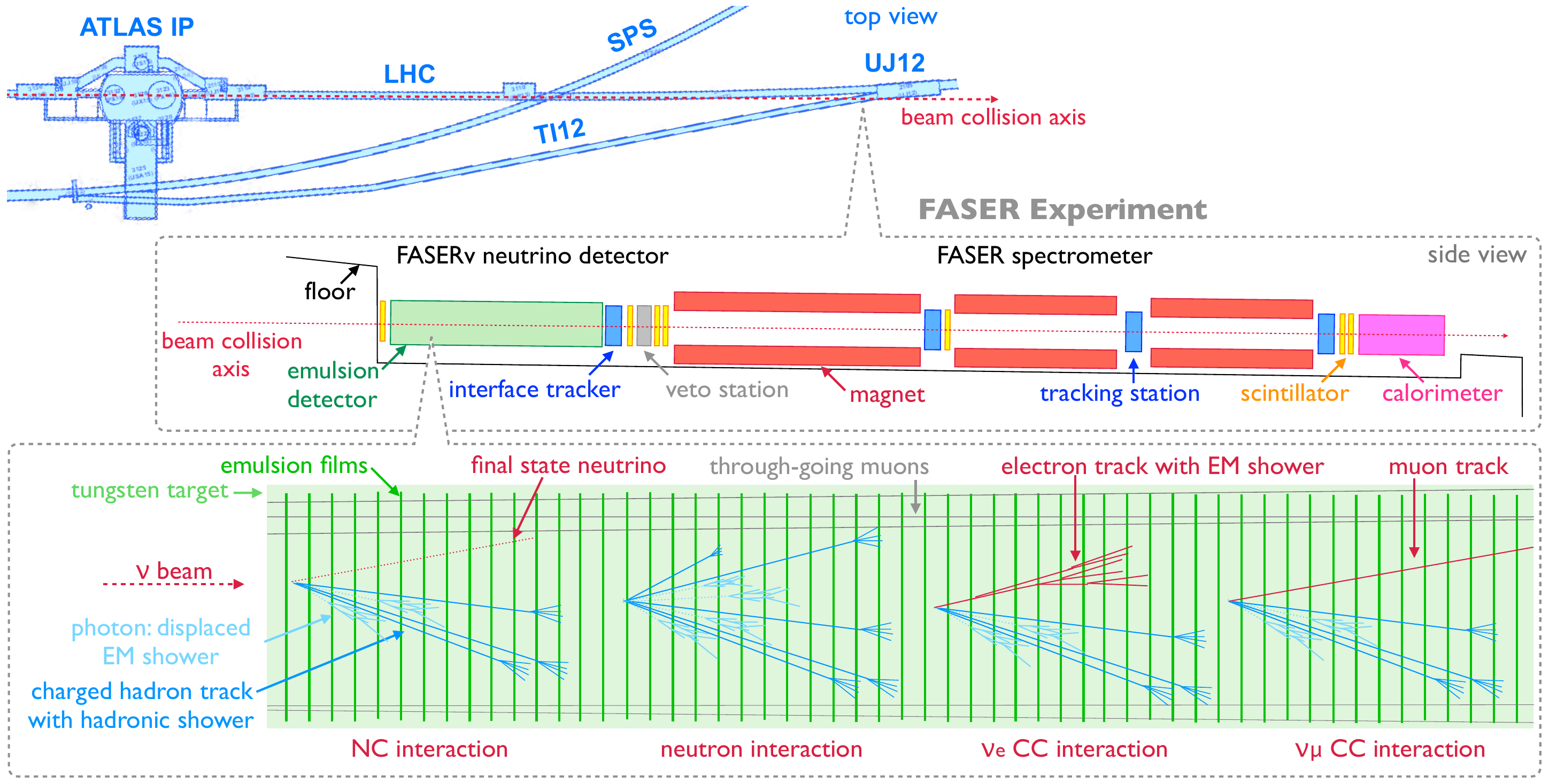}
\caption{Location of the FASER$\nu$ detector and event topology. \textbf{Top:} The FASER experiment is placed about $500$~m downstream of the ATLAS interaction point in the previously unused side tunnel TI12, which connects the SPS with the LHC tunnel. \textbf{Center:} The detector is centered around the beam collision axis where the neutrino flux is maximal. It consists of the FASER$\nu$ emulsion neutrino detector, followed by a magnetized spectrometer and a calorimeter. \textbf{Bottom:} The emulsion detector consists of tungsten plates interleaved with nuclear emulsion films. Both interactions of neutrinos and neutral hadrons lead to the appearance of a \textit{neutral vertex} at which several charged particles emerge. Different types of events can be distinguished based on the event topology, as explained in the text.}
\label{fig:illustration}
\end{figure*}

\section{Neutrino Interactions at \texorpdfstring{\fasernu}{fasernu}}
\label{sec:faserv} 

FASER~\cite{Feng:2017uoz, Ariga:2018zuc, Ariga:2018pin} is a dedicated experiment at the LHC to both search for long-lived particles predicted by models of new physics~\cite{Feng:2017vli, Kling:2018wct, Feng:2018noy, Ariga:2018uku, Berlin:2018jbm, Ariga:2019ufm, Jodlowski:2019ycu, Kling:2020mch, Jodlowski:2020vhr}, and to study interactions of high-energy neutrinos~\cite{Abreu:2019yak, Abreu:2020ddv}. It is located in the far-forward direction, roughly 480 m downstream from the ATLAS interaction point (IP). At this location, the highly collimated neutrino beam produced at ATLAS, which is centered around the \textit{beam collision axis}, intersects with the side tunnel TI12, as shown in the upper part of \cref{fig:illustration}. TI12 has previously served as an injector tunnel for LEP but remained unused during the LHC era. To maximize its sensitivity, a trench has been dug into the floor of TI12 such that the FASER apparatus can be aligned with the beam collision axis. 

A schematic layout of the FASER detector is shown in the center part of \cref{fig:illustration}. Located on the front is the FASER$\nu$ neutrino detector. It is followed by the FASER spectrometer, consisting of magnets and three tracking stations. FASER$\nu$ and the FASER spectrometer are connected by an interface tracking station, which allows a combined analysis of the emulsion and electronic detector components. In addition, the interface tracker can be used to time-stamp the event, which allows for a front veto to reject muon-associated background. At the end of FASER is an electromagnetic calorimeter.

The FASER$\nu$ detector consists of emulsion plates that are interleaved with tungsten plates as a target. This configuration permits the reconstruction of tracks of charged particles passing through the detector with a sub-$\micm$ spatial resolution~\cite{Nakamura:2006xs}. This allows observation of the event topology as shown in the lower part of \cref{fig:illustration}. 

Both neutrino and neutral hadron interactions are expected to produce several hadronic particles forming a collimated jet. This leads to a characteristic \textit{neutral vertex} signature, with several outgoing tracks but no incoming track, that can be searched for. While most neutral hadrons escape undetected, charged hadrons will leave tracks and interact on a length scale of $\lambda_\text{int} \sim 10~\cm$, initiating a hadronic shower. Neutral pions promptly decay into photons, which can be identified by their displaced electromagnetic showers. These showers typically occur within a radiation length $X_0 \sim 3.5~\mm$ in tungsten and point back to the neutral vertex.

It is further possible to distinguish different event types based on their topologies. CC neutrino interaction events contain an energetic charged lepton. While muons can be identified from tracks that do not interact further downstream in the detector, electrons lead to electromagnetic showers that emerge from a track connected to the neutral vertex. NC interactions contain a neutrino in the final state, which escapes undetected and is expected to recoil against the hadronic activity, but no charged leptons. In contrast, neutral hadron interactions lead to a more uniform distribution of hadrons. 

The high spatial resolution of emulsion detectors allows for the use of multiple Coulomb scatterings to estimate the momenta of charged tracks passing through the detector. Momentum measurements of final state charged particles can be used alongside other topological observables to estimate the energy of neutrinos. As shown in Ref.~\cite{Abreu:2019yak}, an energy resolution of about 30\% can be achieved for CC neutrino interactions, while results for NC neutrino interactions are presented in this study. 

\section{Simulation}
\label{sec:simulation} 

The physics signal considered in this paper is NC neutrino scattering. While all flavors of neutrinos can contribute to the signal, the majority of neutrinos passing through FASER$\nu$ are muon neutrinos, supplemented by a sub-leading component of electron neutrinos. In this study, we use the fluxes and energy spectra of neutrinos passing through FASER$\nu$ obtained in Ref.~\cite{Abreu:2019yak}. There it was found that muon neutrinos originate mainly from charged pion and kaon decays, electron neutrinos are primarily produced in neutral kaon, hyperon and $D$-meson decays, and tau neutrinos mainly come from $D_s$ meson decay. All three neutrino flavors have spectra extending over a broad energy range with average energies between $600~\gev$ and $1~\tev$. 

The main background to neutrino NC events comes from high-energy neutral hadrons interacting with the detector. These neutral hadrons are produced by muons striking the tungsten within FASER$\nu$ or the rock in front of it. The flux and energy spectra of muons have been estimated using \texttt{Fluka}, and are presented in the FASER technical proposal~\cite{Ariga:2018pin}. The expected muon rate is about $2 \cdot 10^4~\fb/\cm^2$, which has been validated with \textit{in-situ} measurements. This corresponds to roughly $2\cdot 10^9$ muons passing through FASER$\nu$ during Run 3 of the LHC with a nominal integrated luminosity of $150~\ifb$. Positively charged muons have a softer energy spectrum than negatively charged muons and produce much fewer neutral hadrons, so in what follows we neglect them.

Using these results, we then estimate the rate and energy spectra of neutral hadrons. Using \texttt{Fluka}~\cite{Ferrari:898301, Bohlen:2014buj}, we simulate muons striking a $25~\cm \times 25~\cm \times 1~\m$ block of tungsten, preceded by a large volume of rock. The nuclear interaction length is approximately $\lambda_\text{int} = 10~\cm$, so nearly all the neutral hadrons produced within FASER$\nu$ interact. We find that the number of neutral hadrons at each energy is roughly independent of the longitudinal position within the detector. We obtain the spectra of hadrons interacting with the FASER$\nu$ detector, which is shown in \cref{fig:nhspec}. We can see that the neutral hadron flux is dominated by neutral kaons, followed by neutrons. Neutral hadrons are also produced by neutrino NC (and CC) events themselves, but these are a subdominant contribution to the total flux. 

\begin{figure}[t]
\includegraphics[width=0.49\textwidth]{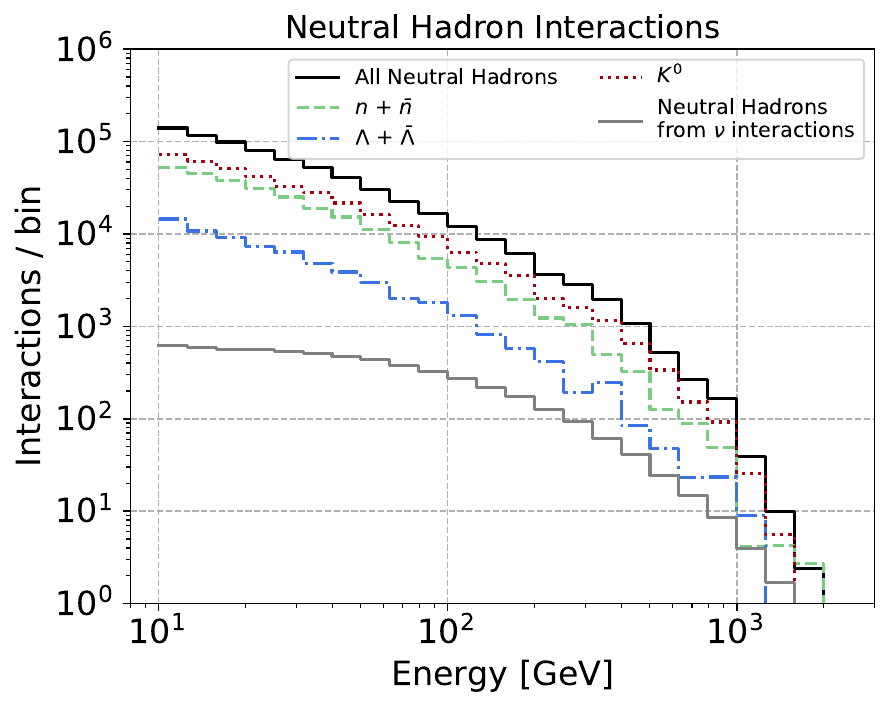}
\caption{Expected energy spectrum of neutral hadrons interacting within the FASER$\nu$ detector during LHC Run~3 with $150~\ifb$ of luminosity.
}
\label{fig:nhspec}
\end{figure}

With these fluxes, we simulate the interactions of both neutrinos and neutral hadrons with \texttt{Pythia~8}~\cite{Sjostrand:2006za, Sjostrand:2014zea} using the \texttt{Monash} tune~\cite{Skands:2014pea}. We use the nuclear parton distribution function \texttt{nCTEQ15} for tungsten~\cite{Kovarik:2015cma, Kusina:2015vfa}. Further nuclear effects in neutral hadron collisions are included with \texttt{Pythia}'s heavy ion module. For simplicity, all neutral hadrons have been simulated as neutrons. We have checked that different types of neutral hadrons forming our background produce similar signatures in the FASER$\nu$ detector, as discussed in the Appendix.

To obtain a first understanding of the uncertainty associated with the simulation, we also simulate neutral hadron collisions with tungsten using \texttt{EPOS-LHC}~\cite{Pierog:2013ria} and \texttt{QGSJET-II-04}~\cite{Ostapchenko:2010vb} as implemented in \texttt{CRMC}~\cite{CRMC}. In addition, we generate neutrino interactions using \texttt{GENIE}~\cite{Andreopoulos:2009rq, Andreopoulos:2015wxa}, following the settings presented in Ref.~\cite{Abreu:2019yak}. While \texttt{GENIE}'s simulation of deep inelastic scattering (DIS) events is based on \texttt{Pythia~6}, it also takes into account final state interaction effects. We find that the different simulations are in good agreement. However, we note that dedicated future efforts are needed to further validate and improve the simulation such that the associated uncertainties can be quantified and reduced. A detailed comparison between different simulators is presented in \cref{sec:appendix-gnrts}. 

In order to simulate the detector response, we perform a phenomenological detector simulation as follows. First, we choose a location for the primary interaction vertex within the detector from a random uniform distribution. In the next step, observable final states, such as charged tracks, photons, and electrons, are identified. Unobservable final state particles, such as neutrinos and neutral hadrons, are rejected. We also remove soft particles with energies below $300~\mev$. We then assign a momentum to each observed particle using its energy and direction. While emulsion detectors can measure the directions of final state particles very accurately, we smear the energies (obtained either via the electromagnetic shower for electrons and photons or from multiple Coulomb scattering for tracks) according to the results obtained in Ref.~\cite{Abreu:2019yak}: We use a Gaussian smearing with width $\sigma_E/E=50\%$ for showers and $\sigma_E/E=46\%$ for charged particles. Finally, charged hadron tracks will often undergo secondary interactions, which will later be used to distinguish them from muons. The distance between primary and secondary interactions is sampled according to its exponential probability distribution. Both muons and charged hadrons that are produced without interacting again inside the detector are marked as muon candidates.

\section{Analysis}
\label{sec:analysis} 

As described previously, emulsion detectors such as FASER$\nu$ are able to identify neutral vertices and also to record associated kinematic and topological features. In this section, we will present a neural network-based analysis to separate the NC neutrino interaction signal from the neutral hadron background and estimate the energy of the incoming neutrino. In the following, we will define a set of observables characterizing the interactions in \cref{sec:observables}, and then discuss their use in signal identification (\cref{sec:signal_id}) and neutrino energy estimation (\cref{sec:energy_estimation}).

\subsection{Event observables}
\label{sec:observables}

\begin{figure*}[th!]
\includegraphics[width=1.00\textwidth]{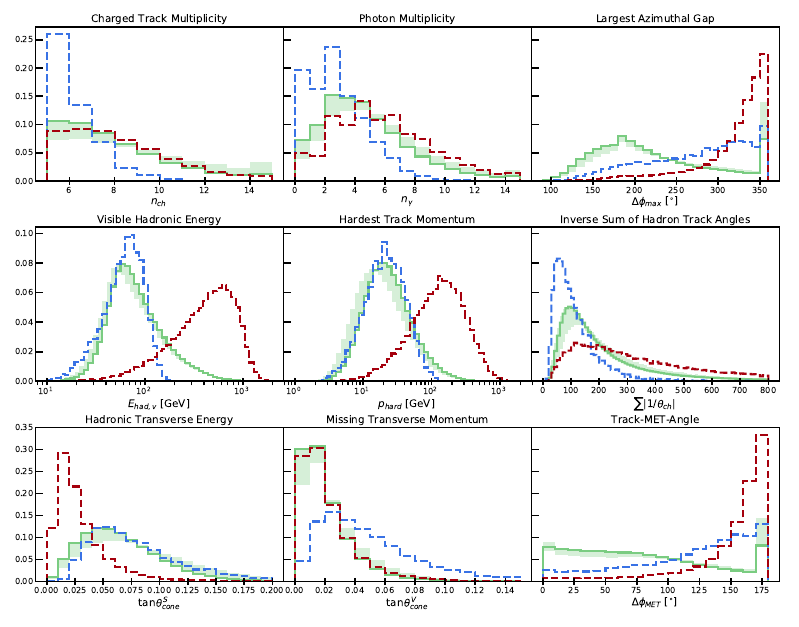}
\caption{Normalized kinematic distributions for the observables defined in \cref{sec:observables}. The dashed lines show the distributions obtained with \texttt{Pythia~8} for the NC neutrino interaction signal at incoming neutrino energies of $E_\nu=100~\gev$ (blue) and $E_\nu=1~\tev$ (red). The solid green lines correspond to the distributions for the neutral hadron interactions simulated with \texttt{Pythia~8} for the expected energy spectrum presented in \cref{fig:nhspec}. The shaded region shows the range of predictions for the background distributions obtained from different generators: \texttt{Pythia~8}, \texttt{EPOS-LHC} and \texttt{QGSJET-II-04}. 
}
\label{fig:obs}
\end{figure*}

Due to its high spatial resolution, FASER$\nu$ is able to precisely measure geometric variables, such as the multiplicities of tracks and photons and the directions, as well as to estimate kinematic quantities, such as charged particle momenta and energies of electromagnetic showers. We will now use these features to define a set of observables, which will subsequently be used as input for a multivariate analysis. Note that we will only consider tracks and showers with energy $E>1~\gev$ and an angle of $\theta < 45^\circ$ with respect to the incoming particle direction for the construction of these observables. This is to reduce the dependence of soft hadronic physics effects that might not be modeled accurately by MC simulators. 

\begin{itemize}
    \item The \textit{charged track multiplicity} ($n_\text{ch}$) is the number of charged tracks originating from the neutral vertex. It is related to the hadronic energy in the event: $n_\text{ch} \sim \log(E_\text{had})$~\cite{Grosse:2009kz}. Events considered in this study have $n_\text{ch}\geq 5$ charged tracks as required by the neutral vertex selection~\cite{Abreu:2019yak}.
    \item Similarly, the \textit{photon multiplicity} ($n_\gamma$) measures the number of identified photon-initiated electromagnetic showers that have been associated with the neutral vertex. The observed photons mainly originate from the prompt decays of neutral pions, making them a proxy for the pion multiplicity $n_\gamma \sim 2 n_{\pi^0}$ and the hadronic energy $n_\text{ch} \sim \log(E_\text{had})$. 
    \item The \textit{visible hadronic energy} ($E_\text{had,v}$) can be measured as the sum of reconstructed energies of visible particles, $E_\text{had,v} = \sum E_\text{ch} + \sum E_\gamma$, which includes both charged tracks and photons (mainly from neutral pion decay). It is proportional to the true hadronic energy, $E_\text{had,v} \sim E_\text{had}$, which also includes additional long-lived neutral hadrons. 
    \item Additionally, we consider the \textit{momentum of the hardest track} ($p_\text{hard}$). It is closely related to the hadronic energy of the event $p_\text{hard} \sim E_\text{had}$.
    \item Another observable is the \textit{inverse sum of track angles} ($\sum |1 / \theta_\text{ch}|$), where $\tan\theta_\text{ch} = p_T/p_z$ is the slope of the individual tracks. More energetic events have hadron tracks with smaller angles and hence a larger value for this observable, $\sum |1/\theta_\text{ch}| \sim E_\text{had}$.
    \item The \textit{scalar cone angle} ($\tan\theta_\text{cone}^S$) is defined as the scalar sum of the momentum-weighted track angles, $\tan\theta_\text{cone}^S \!=\! \sum p_i \tan\theta_i / \sum p_i = \sum p_{T,i}/\sum p_i $. It is proportional to the \textit{hadronic transverse energy} ($H_T$) of the event, $\tan\theta_\text{cone}^S \sim H_{T} / E_\text{had}$.
    \item Additionally, the \textit{vector cone angle} ($\tan\theta_\text{cone}^V$) is defined as the vector sum of the individual track angles weighted by their momenta, with two components corresponding to the $x$ and $y$ directions. Using the tracks' transverse momenta, $\vec{p}_{T,i}$, it can be written as $\tan\theta_\text{cone}^V = \sum \vec{p}_{T,i} / \sum p_i$. It is proportional to the \textit{missing transverse momentum} ($\slashed{\vec{p}}_T$) of the event, $\tan\theta_\text{cone}^V \sim \slashed{\vec{p}}_T / E_\text{had}$. Larger values for the missing transverse energy , $|\slashed{\vec{p}}_T|$, are expected for NC neutrino events, in which the final state neutrino will carry away a sizable fraction of the incoming neutrino energy.
    \item The \textit{largest azimuthal gap} ($\Delta \phi_\text{max}$) is defined as the largest difference in azimuthal angle between two neighboring visible particles (charged tracks and photons) whose energy is $> 0.1 E_\text{had,v}$. This angle will be large for events in which a neutrino recoils against all of the hadronic activity  ($\Delta \phi_\text{max}>\pi$), and small for background events without a neutrino.
    \item Similarly, the \textit{track MET angle} ($\Delta \phi_\text{MET}$) is the azimuthal angle between the reconstructed missing transverse momentum, $\slashed{\vec{p}}_T$, and the nearest visible particle with energy $> 0.1 E_\text{had,v}$. This angle should be large for NC neutrino interaction events ($\Delta \phi_\text{MET}>\pi/2$), and small for neutral hadron events.
\end{itemize}

In \cref{fig:obs}, we show the kinematic distributions for these observables. The dashed lines correspond to NC neutrino interactions with a neutrino energy of $E_\nu=100~\gev$ (blue) and $1~\tev$ (red). The solid green lines show the distributions for the neutral hadron background with the energy spectrum shown in \cref{fig:nhspec}. The shaded band corresponds to the range of predictions obtained from different generators, serving as a rough estimate of the background simulation uncertainty. We can see that the generator predictions are generally consistent and that the differences between different simulators are mild. 

The most striking differences between the signal and background can be observed in the azimuthal angle features $\Delta \phi_\text{MET}$ and $\Delta \phi_\text{max}$. Large values for these observables are caused by the presence of a neutrino in the final state recoiling against hadrons. In contrast, neutral hadron interactions are expected to produce a more uniform angular distribution of charged tracks, leading to smaller values for these angles. Note that when calculating these observables, we only consider visible particles whose energy is larger than 10\% of the visible hadronic energy. If only one track passes this energy threshold, $\Delta\phi_\text{max}=360^\circ$. We also note that most of the backgrounds at FASER$\nu$ lie in the low energy range, $E \lesssim$ few $100~\gev$. This explains why the background distributions are often similar to the $E_\nu=100~\gev$ curve but also implies that the energy content can be used to distinguish the neutral hadron background from NC signal with typical energies $E_\nu \sim \tev$.

\begin{figure*}[t!]
\centering

\includegraphics[width=0.99\textwidth]{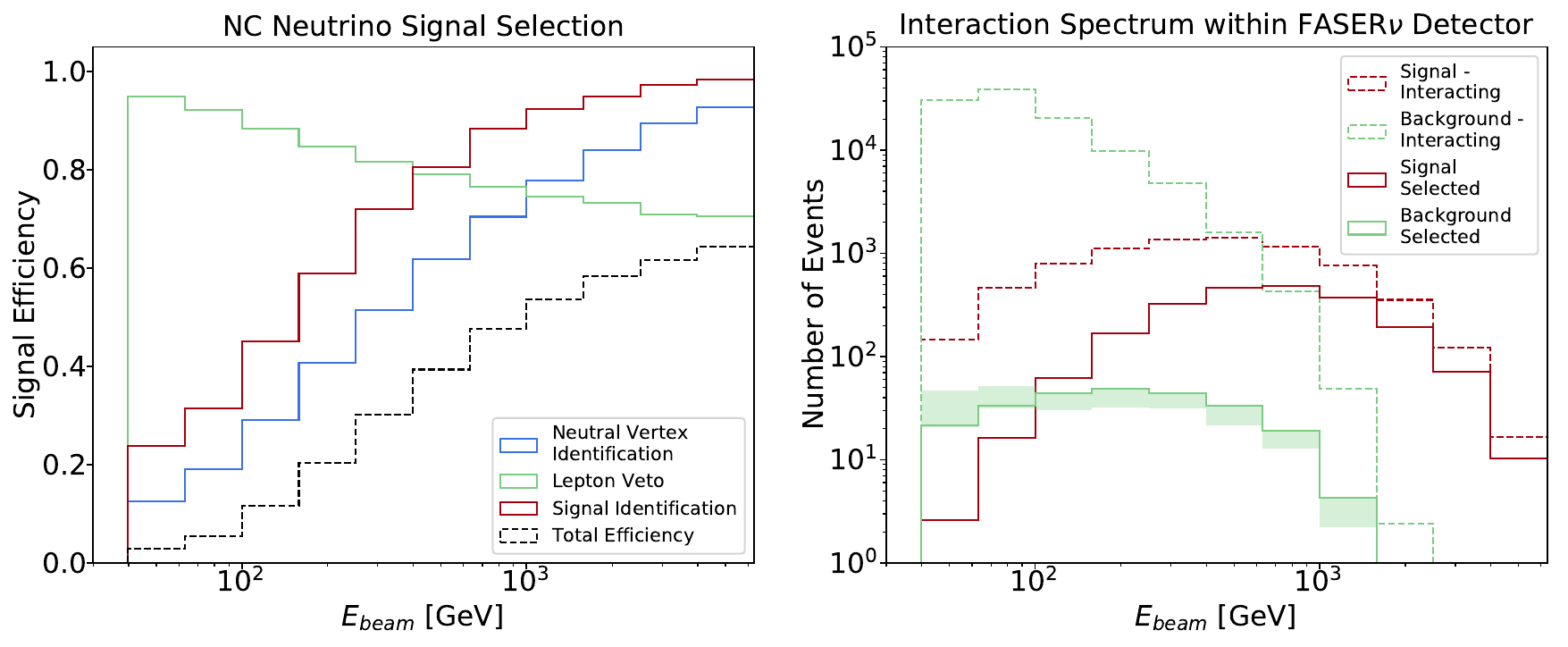}
\caption{\textbf{Left:} Signal selection efficiencies as a function of beam energy. Each line indicates the fraction of events passing the following criteria sequentially: i) neutral vertex identification (blue) requiring $\geq$ 5 charged tracks, ii) lepton veto (green) requiring no electron candidate and no non-interacting charged track, and iii) signal identification (red) as performed by the NN classifier. The dashed black line shows the combined efficiency. 
\textbf{Right:} The energy spectra of particles interacting within the FASER$\nu$ detector. We show the expected numbers of neutral hadron interactions (green) and NC neutrino interactions (red) with the FASER$\nu$ detector during LHC Run~3 as dashed lines. The solid lines show the spectra for events passing the signal selection (including neutral vertex identification, lepton veto and signal identification). The uncertainty associated to the background generation is shown as a shaded band. 
}
\label{fig:classification}
\end{figure*}

Comparing the distributions for the two considered beam energies, $E_\nu=100~\gev$ and $1~\tev$, we can see that all observables are sensitive to the incoming neutrino energy. Not surprisingly, the largest differences are observed for track momentum-based observables: the visible hadronic energy, $E_\text{had,v}$, and the momentum of the hardest track $p_\text{hard}$. However, complementary information is also provided by the other observables, motivating the multivariate analysis to obtain more robust results. 

Note that here we assume that all incoming particles are moving parallel to the beam collision axis. In reality, the incoming neutrinos have an angular spread of $\theta \sim 0.5~\mrad$, corresponding to the angular size of the detector. In addition, neutral hadrons, which are the result of scattering events occurring close to FASER$\nu$, will also have a small angle with respect to the beam axis of $\theta \lesssim 10~\mrad$ for energies $E>100~\gev$, as shown in Ref.~\cite{Abreu:2019yak}. These incoming beam angles are smaller than typical values of $\tan\theta_\text{cone}^{S,V}$. However, small transverse momenta of incoming neutral hadrons and neutrinos can potentially distort the observable distributions, and should therefore be taken into account in a full experimental analysis. 

\subsection{Signal identification}
\label{sec:signal_id}

Let us now turn to the selection of NC neutrino interaction events. We first require the presence of a neutral vertex. Following Ref.~\cite{Abreu:2019yak}, we demand the presence of $\geq 5$ charged tracks with momentum $p >1~\gev$ and slope $\theta < 45^\circ$ emerging from the vertex. The resulting neutral vertex identification efficiency for NC neutrino interactions is shown as the blue line in the left panel of \cref{fig:classification}. It is strongly suppressed at lower energies due to the typically lower charged particle multiplicity, but attains values of $> 80\%$ for neutrino energies $E_\nu > 1~\tev$.

In the second step, we veto all events containing a charged lepton candidate in the final state. Here each charged track with more than $5\%$ of the event's visible hadronic energy that leaves the detector before interacting is considered a muon candidate. While designed to effectively eliminate the CC neutrino interaction background, it also reduces the acceptance rate for the NC neutrino interaction signal, especially for interactions occurring toward the end of the detector. The efficiency of NC events to pass the charged lepton veto is shown as the green line in the left panel of \cref{fig:classification}. At TeV energies, the efficiency is about 80\%. The efficiency increases toward lower energies, mainly due to the typically lower multiplicity of charged tracks that could be potentially misidentified as muons. We assume that the fraction of CC events passing the lepton veto is negligible.

\begin{figure*}[th!]
\includegraphics[width=0.99\textwidth]{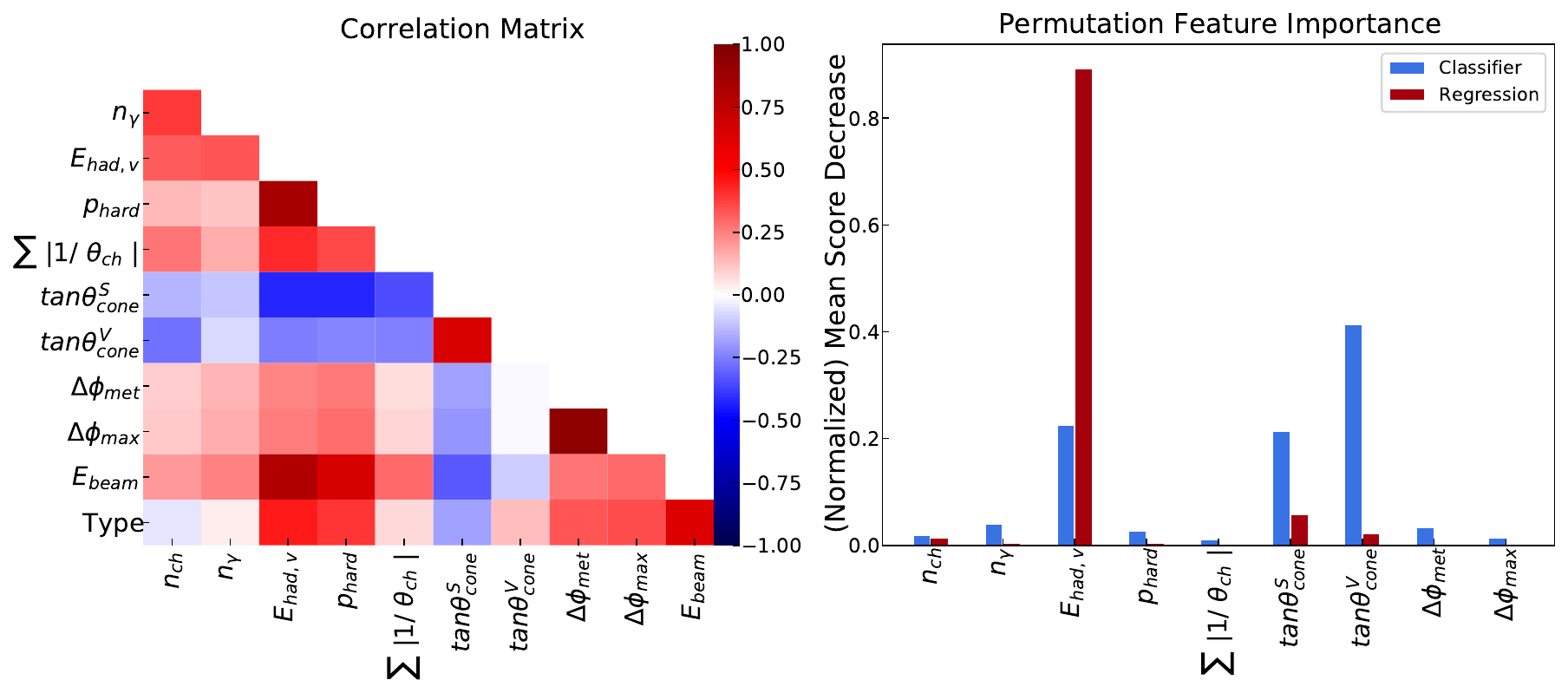}
\caption{\textbf{Left:} Correlation matrix showing the linear relationship between observables presented in \cref{sec:observables}, the incoming particle energy ($E_\text{beam}$) and the event type (1 for NC neutrino interactions, 0 for neutral hadrons).
\textbf{Right:} Permutation feature importance (the normalized mean score decrease for each of the observables) for the signal identification classifier (blue) and neutrino energy estimator (red) network. We use accuracy as the score metric for the classifier network and mean average percent error for the estimator. Scores decreases are normalized so that they sum to 1.}
\label{fig:importance}
\end{figure*}

After removing CC neutrino interactions, we are left with the NC neutrino interaction signal and neutral hadron interaction backgrounds. In this work, we will separate the two samples using a neural network classifier, which uses the observables introduced in the previous section as input. 
 
We simulate 100 times the expected Run 3 event rate for both the NC neutrino interaction signal and the neutral hadron interaction background using \texttt{Pythia~8}. We then train a neural network in \texttt{Keras}~\cite{keras} to classify the event type as either signal or background. We use a fully connected neural network with three hidden layers of 64 units and a sigmoid activation function, minimizing the binary cross-entropy loss by training with the \texttt{Adam} optimizer over 20 epochs. Our training employed a batch size of 256, a constant learning rate of $10^{-3}$, and early stopping to avoid overtraining. These hyperparameters are the result of a coarse manual scan, and we did not perform an exhaustive optimization. It is likely that performance could be further improved with additional tuning.

The resulting signal identification efficiency is shown as the red line in the left panel of \cref{fig:classification}. The combined efficiency of vertex identification, lepton veto and signal identification is shown as the dashed black line. It is approximately 50\% at TeV energies, but significantly reduced at lower energies, mainly due to the low neutral vertex detection efficiency. 

In the right panel of \cref{fig:classification} we show the energy spectra of NC neutrino interactions (red) and neutral hadron interactions (green). The dashed lines correspond to all interactions occurring within the detector. After applying all signal selection criteria, the event rates drop to the solid lines. We can see that the neural network classifier is able to identify the signal and sufficiently reduce the background. While the classification network was trained with \texttt{Pythia~8}, we have also tested it on data produced with \texttt{EPOS-LHC} and \texttt{QGSJET-II-04}, assuming the same incoming spectrum of neutral hadrons. The result is shown as a shaded band around the background line. We can see that uncertainties arising from the different simulation of neutral hadron interactions are small, and do not change the background rates significantly. 

Before moving on, we further discuss the trained classifier network to understand which observables are most relevant for the signal identification. In the left panel of \cref{fig:importance}, we show the correlation matrix between different observables, the beam energy and the event type. Dark shaded bins correspond to stronger linear correlations, either positive (red) or negative (blue). We can see that the event type is most strongly correlated with the visible hadronic energy and the momentum of the hardest track. This is expected as the incoming neutrinos which interact with the detector tend to be harder than the neutral hadrons. We also see that  the more energy associated with an event (larger $E_\text{beam}$, $E_\text{had,v}$, $p_\text{hard}$), the more tightly collimated its reaction products are (smaller cone angles $\theta_\text{cone}$, larger azimuthal angles $\Delta \phi$). The full network, of course, has the ability to learn non-linear relationships.

In the right panel of \cref{fig:importance}, we show another common tool to analyse the network's performance: the \textit{permutation feature importance}. It is obtained by randomly shuffling the values of one observable (say $n_\text{ch}$) between different events and recording the degradation in the final score obtained by the network. For the classifier network, the accuracy is taken as the score. Large decreases in the accuracy when randomizing a given observable indicate that the observable is important for network performance. The blue bars show the results for the event classification network. We can see that the most important variables for classification are the hadronic energy and the cone angles, or, equivalently, $\slashed{\vec{p}}_T$ and $H_T$. By contrast, when two observables provide the same information to the network, the permutation importance of each is low. This happens with the angular variables $\Delta \phi_\text{max}$ and $\Delta \phi_\text{MET}$: while they are clearly correlated with the event type, each variable gives the same information, so removing one alone does not significantly harm the network performance.

\begin{figure*}[th!]
\includegraphics[trim=0 5 0 5, clip, width=0.99\textwidth]{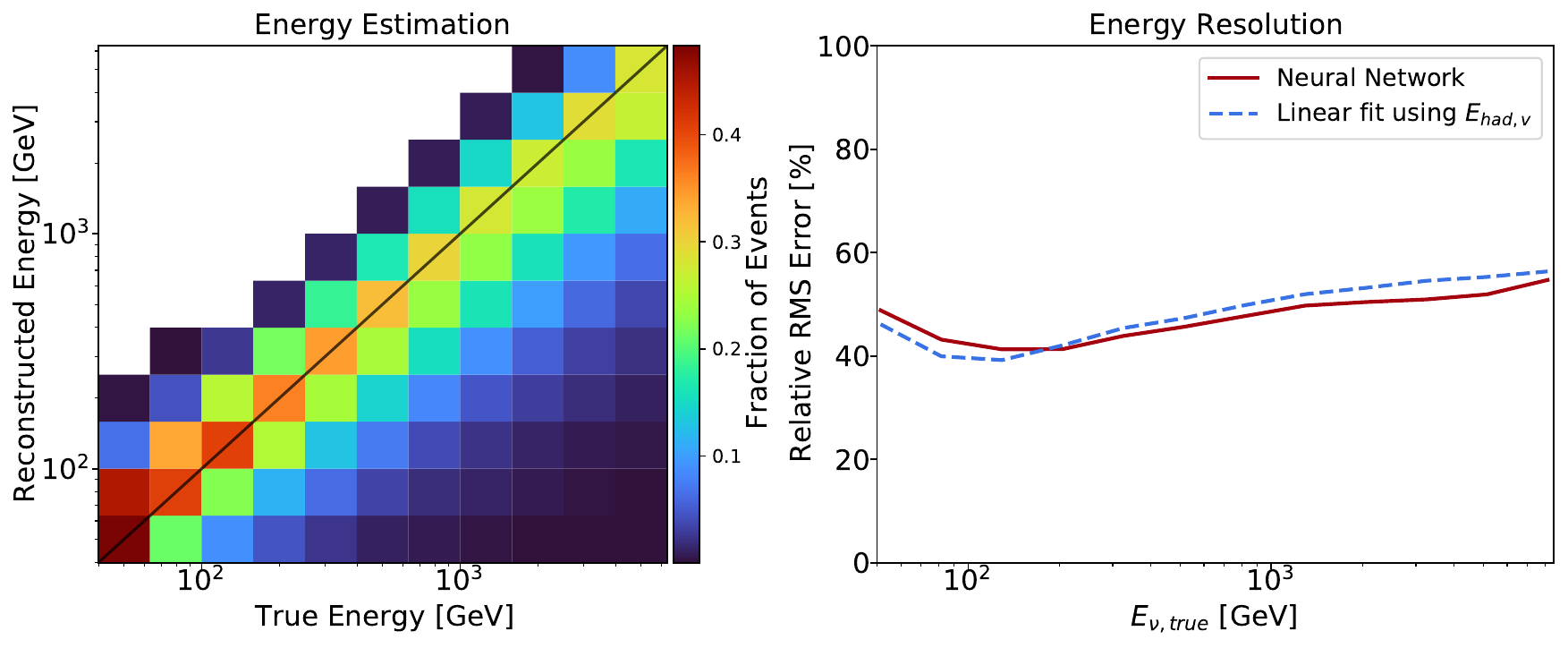}
\caption{\textbf{Left:} Neutrino energy reconstruction for NC neutrino interaction events obtained by a neural network-based multivariate analysis using the observables defined in \cref{sec:observables}. \textbf{Right:} Relative RMS energy resolution using the neural network-based multivariate analysis (solid) and only the hadronic energy of the events (dashed).}
\label{fig:energyreco}
\end{figure*}

\subsection{Neutrino energy estimation}
\label{sec:energy_estimation}

Having discussed the selection of NC neutrino interaction events, let us now turn to the estimation of the incoming neutrino's energy. In DIS neutrino interactions at FASER$\nu$, roughly half of the incoming neutrino energy is transferred to the nucleus on average. Since the final state neutrino escapes undetected, the observable hadronic final state is the only handle for energy reconstruction. The absence of an observable lepton results in degraded energy resolution compared to results obtained for CC neutrino interactions in Ref.~\cite{Abreu:2019yak}. 

As we have seen in \cref{sec:observables}, all observables considered are sensitive to the neutrino energy, motivating the use of multivariate energy estimation. We simulate $6\cdot 10^5$ NC neutrino interaction events that are uniformly distributed in $\log E_\nu$ using \texttt{Pythia~8}, and train a neural network to minimize the mean average percent error between the true and estimated energy. Here we use the same network architecture and hyperparameters as for the classification network, except the final layer has an identity map as its activation function.

The result of the energy estimation is shown in \cref{fig:energyreco}. The left panel shows the correlation between the reconstructed and true energy. With five bins per decade in energy, the leakage of events between bins is mild, indicating that the neutrino energy estimation for neutral current events is indeed possible. The right panel shows the RMS energy resolution
\begin{equation}
\textrm{relative RMS error} =\! \sqrt{\langle \left(E_{\nu,\text{reco}} \!-\! E_{\nu,\text{true}}\right)^2 \!/E_{\nu,\text{true}}^2\rangle} \!
\end{equation}
as a function of energy. We obtain an energy resolution of about 50\%.

As for the signal identification network, let us study which observables are most important for the energy estimation. In the left panel of \cref{fig:importance}, we see that the incoming particle energy $E_\text{beam}$ is particularly well-correlated with the visible hadronic energy, though there is also a clear relationship with $p_\text{hard}$. In the permutation importance study for the energy estimation network shown by the red bars in the right panel, we use the increase in mean average percent error to quantify the impact of randomly permuting the values of one observable among events. $E_\text{had,v}$ is by far the dominant observable, suggesting that our network has learned the strong correlation between the visible hadronic energy and that of the incoming neutrino, and is relying heavily on the former to estimate the latter. This dependence arises regardless of the correlation between $E_\text{had,v}$ and the momentum of the hardest track, which is perhaps not surprising as $p_\text{hard}$ is not as directly correlated with the neutrino energy. While the fraction of the neutrino energy that is transferred to the nucleus has an almost uniform distribution, we find that the visible hadronic energy still serves as an excellent proxy for the energy of the incoming neutrino. Motivated by this, we also show in the right panel of \cref{fig:energyreco} the energy resolution that can be obtained by a linear fit to the visible hadronic energy. The almost similar performance demonstrates the clear importance of the visible hadronic energy to neutrino energy estimation.

\section{Results and Interpretation}
\label{sec:results} 

\subsection{NC cross section measurements}
\label{sec:nc_xs}

\begin{figure*}[th!]
\includegraphics[trim=0 5 0 5, clip, width=0.99\textwidth]{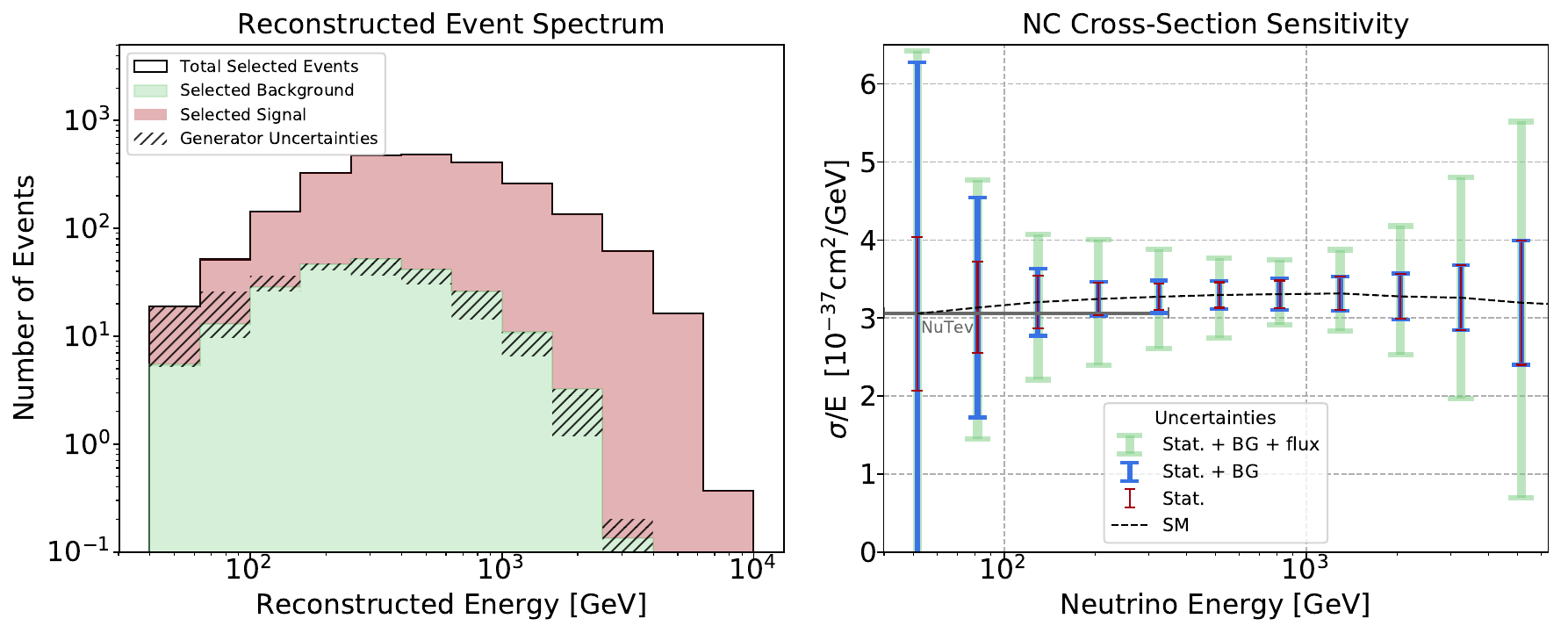}
\caption{
\textbf{Left:} Stacked histogram of events passing the event signal selection described in \cref{sec:signal_id} as function of the reconstructed energy for LHC Run 3 with $150~\ifb$ integrated luminosity. The red and green shaded regions show the NC neutrino interaction signal and the neutral hadron interaction background, respectively. The hatched region indicates the uncertainty arising from the simulation of neutral hadron interactions, corresponding to the range of predictions obtained by three different generators. 
\textbf{Right:} FASER$\nu$'s estimated neutrino-tungsten NC cross section sensitivity. Existing constraints are shown in gray. The black dashed curve is the theoretical prediction for the DIS cross section, averaged over neutrinos and anti-neutrinos, per tungsten nucleus. The inner red error bars correspond to statistical uncertainties, the blue error bars additional take into account uncertainties associated with the simulation of the background, and the outer green error bars show the combined uncertainties with the neutrino production rate (which corresponds to the range of predictions obtained from different MC generators as obtained in Ref.~\cite{Abreu:2019yak}).} 
\label{fig:eventspectrum}
\end{figure*}

With the analysis described in the previous section, we proceed to discuss FASER$\nu$'s expected physics sensitivity. In the left panel of \cref{fig:eventspectrum}, we show the expected number of NC neutrino signal (red) and neutral hadron background (green) events simulated with \texttt{Pythia~8} passing the event selection as a function of the reconstructed energy. As noted before, the signal dominates over the background for energies above about $100~\gev$ and reaches a signal to background ratio $\gtrsim 100$ for energies above $1~\tev$. The hatched region shows the background simulation uncertainty, corresponding to the range of predictions obtained from three different event generators to simulate neutral hadron interactions using the same neutral hadron flux and energy spectrum. 

Assuming no new physics contribution to neutrino production and propagation, the observed energy spectrum at FASER$\nu$ can be used to measure the NC neutrino interaction cross section. We show FASER$\nu$'s expected sensitivity to   constraining the NC neutrino interaction cross sections with a tungsten nucleus in the right panel of \cref{fig:eventspectrum}. The black dashed line shows the SM prediction for the cross section, flux-weighted over neutrinos and anti-neutrinos. We also show the NuTeV neutrino-quark neutral current strength measurement~\cite{Zeller:2002he} in gray, which had superior precision, $\mathcal{O}\left(1\% \right)$ with error bars that are too small to be visible,  but used neutrinos that were less energetic than the bulk of the FASER$\nu$ neutrino spectrum. 

Several sources of uncertainties contribute to the measurement. In the following, we discuss these uncertainties and how they could be reduced in a full experimental analysis.

\begin{description}

\item[Statistical uncertainty] During LHC Run 3 with a nominal integrated luminosity of $150~\ifb$, FASER$\nu$ will collect roughly 7000 NC neutrino interactions. The corresponding statistical uncertainties in each energy bin are shown as thin red error bars. 

\item[Neutrino flux]
The neutrino flux uncertainty is associated with the modeling of forward particle production, which is mostly governed by non-perturbative physics and typically described by hadronic interaction models. Here we use the neutrino flux obtained in Ref.~\cite{Abreu:2019yak}, where the range of predictions obtained by different hadronic interactions models was used to estimate the neutrino flux uncertainties. We note that more efforts are needed, and indeed already ongoing, to both quantify and reduce these uncertainties. We show the neutrino flux uncertainty as the green error bars in Fig.\ref{fig:eventspectrum}, and note that this is expected to be a dominating source of uncertainty.  

In extracting limits on new physics, the flux uncertainty can be mitigated by taking the ratio of neutral current to charged current events. This technique has been used by previous experiments~\cite{Holder:1977gq, Allaby:1987vr, McFarland:1997wx, Zeller:2001hh} to measure the weak mixing angle.

\item[Signal simulation] As outlined in Ref.~\cite{Abreu:2020ddv}, there are a variety of uncertainties effecting the signal simulation, including (i) nuclear effects (such nuclear shadowing and anti-shadowing and EMC effect), (ii) the hadronization of final state partons, and (iii) the modeling of final state interactions in the tungsten target nuclei. Currently, there is no neutrino interaction generator that targets this high-energy DIS regime. While recent efforts on nuclear PDFs allow one to describe nuclear effects and their uncertainties~\cite{Eskola:2016oht, Kovarik:2015cma, Kusina:2015vfa, AbdulKhalek:2019mzd, AbdulKhalek:2020yuc}, more dedicated efforts are needed to tune and improve the modeling of hadronization and final state interactions in existing generators and to quantify the uncertainties. In principle, data from previous neutrino experiments, such as DONuT or CHORUS, as well as FASER$\nu$'s CC measurements could be helpful in this regard. 

Uncertainties on the signal simulation will affect the distributions of observables and hence induce uncertainties in all parts of the analysis, including the neutral vertex identification efficiency, the signal identification efficiency, and the energy reconstruction performance. As no reliable estimates of these uncertainties are currently available, we do not attempt to quantify the impact of generator uncertainties on our final results. 

\item[Neutral hadron flux] Analogous to the aforementioned uncertainty on the size of the signal, there are also uncertainties in the numbers of neutral hadrons impinging on FASER$\nu$. The calculation of the neutral hadron flux takes the muon flux in front of FASER$\nu$ as input and relies on the modeling of neutral hadron production from muons interacting with the detector and rock in front of it. The muon flux used in this study was obtained by the CERN STI group using a dedicated \texttt{Fluka} simulation, and it would not be unreasonable to allow for an $\mathcal{O}(1)$ uncertainty on the number of neutral hadrons~\cite{Ariga:2018pin}. Even such a large error, though, is expected to have a small impact on the final cross section uncertainty due to the efficiency of the classification network. The neutral hadron contamination of events that are classified as neutrino interactions is below 10--20\% for energies above $200~\gev$. Furthermore, at Run 3, FASER will directly measure the muon flux and energy spectrum, allowing for reduction of the uncertainty of the input for the neutral hadron calculation. In addition, the number of neutral hadron interactions in FASER$\nu$ can be constrained directly using both measurements of a neutral hadron control sample, as well as charged hadrons which leave clearly visible tracks. 

\item[Background simulation] As shown in \cref{fig:obs}, different generators for neutral hadron interactions produce variations in the distributions of the observables that are used for our analysis. This leads to an uncertainty on the rate of background events passing the event selection, as indicated by the hatched region in the left panel of \cref{fig:eventspectrum}. We have included the resulting uncertainty as blue error bars in the right panel of \cref{fig:eventspectrum}. While this uncertainty dominates the NC neutrino cross section sensitivity at low energies below $100~\gev$, it only mildly affects the measurement at higher energies. These uncertainties can be further improved both using FASER$\nu$ and measurements from dedicated beam dump experiments, such as DsTau~\cite{Aoki:2019jry} and NA61~\cite{Gazdzicki:995681}.

\item[Experimental Uncertainties] While we have incorporated detector effects in our simulation, we do not include experimental uncertainties regarding the detector performance.

\item[Energy estimation] We have estimated the incoming neutrino energy with an error of approximately 50\% for events classified as neutrino neutral current events, as shown in Fig.~\ref{fig:energyreco}. In an experimental analysis, a transfer matrix among the bins could be derived from the network performance. Then, the obtained energy distribution could be unfolded to obtain a better approximation of the incoming neutrino energies. At our level of precision, it is reasonable to assume that this matrix is approximately diagonal given the width of the energy bins, and we do not consider this uncertainty further.

\end{description}

Our results for the neutrino NC cross section are summarized in the right panel of Figure~\ref{fig:eventspectrum}. The most significant source of uncertainty is the neutrino flux at higher energies and the background simulation at energies below $100~\gev$. We note that statistical uncertainties could be reduced with a neutrino detector in the forward region of the HL-LHC, which has a nominal integrated luminosity of $3000~\ifb$. Such a detector could be placed in a future Forward Physics Facility~\cite{Feng:2020fpf} at the High Luminosity LHC.

\subsection{Non-standard interactions}
\label{sec:nsi}

\begin{figure*}[th!]
\centering
\includegraphics[width=0.48\textwidth]{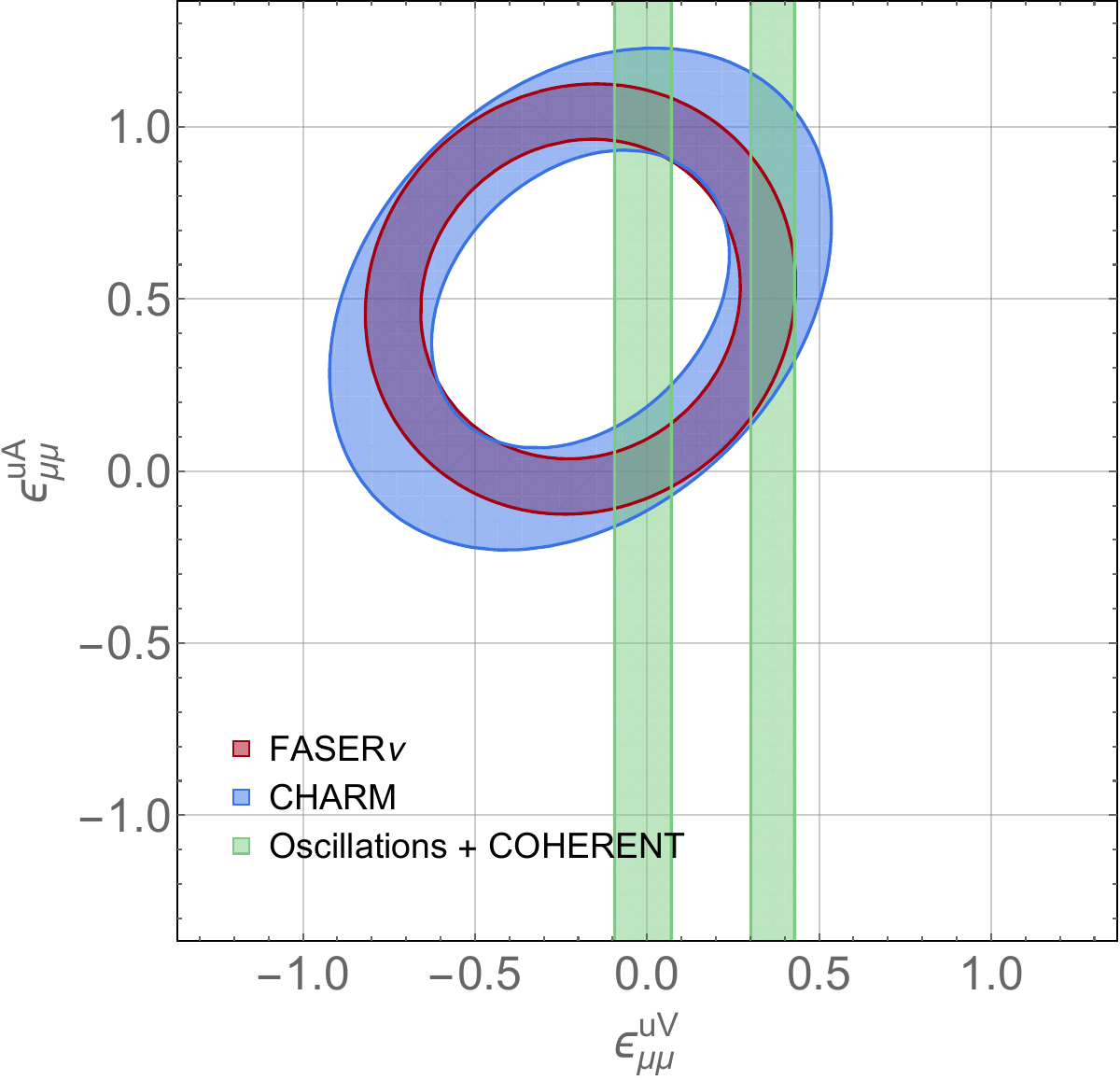}
\hspace{0.3cm}
\includegraphics[width=0.48\textwidth]{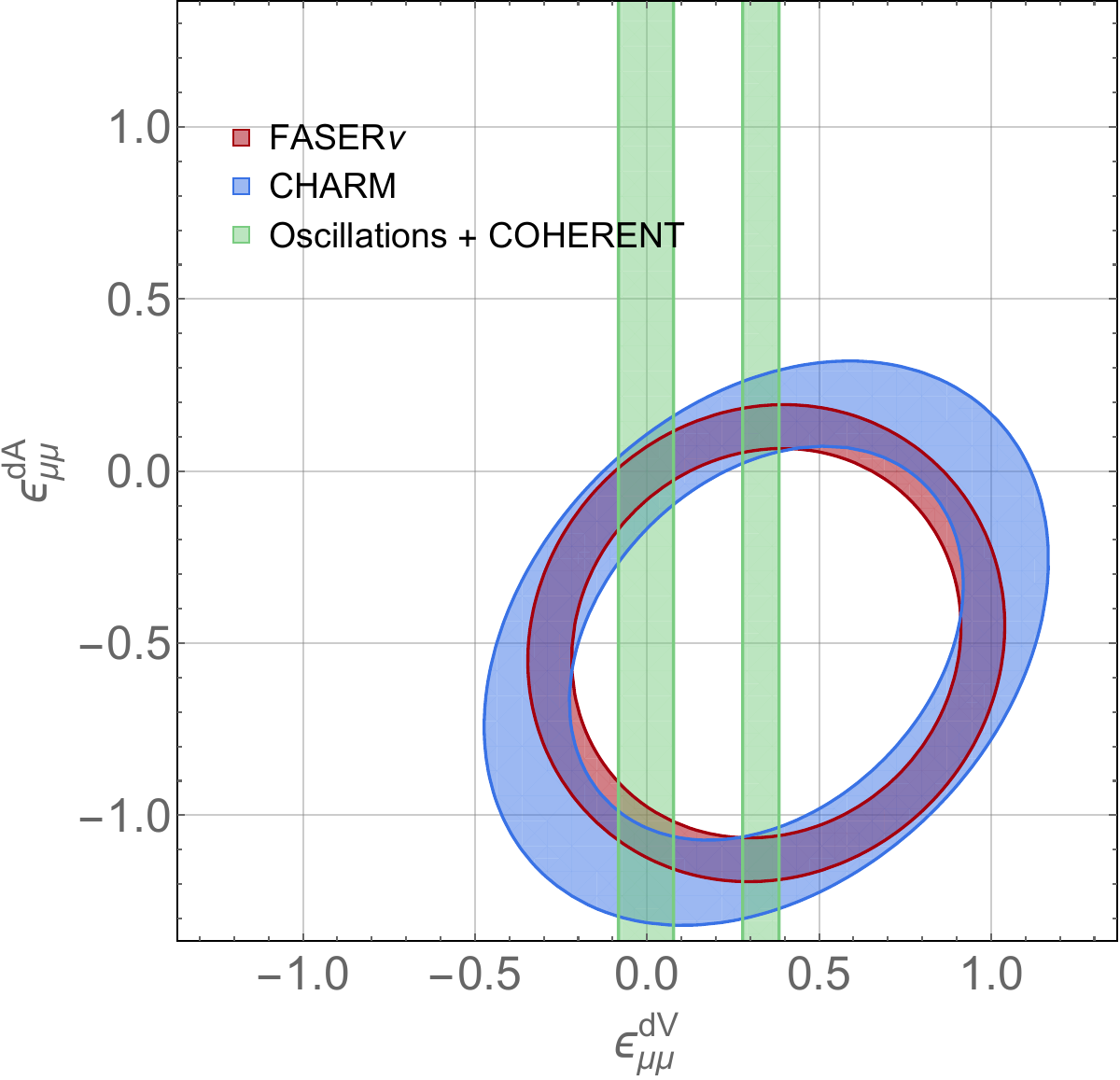}
\caption{ \textbf{Left:} Limits on neutrino NSI involving the up quark. The red ellipse indicates the expected 95\% allowed region by \fasernu, with limits from CHARM~\cite{Allaby:1987vr} (blue) shown for comparison. The one-dimensional allowed region from oscillation and COHERENT~\cite{Coloma:2019mbs} is also shown (green). \textbf{Right:} Same as left plot but for NSI involving the down quark. 
}
\label{fig:nsilimits}
\end{figure*}

The neutrino neutral current cross section can be used to probe new interactions between neutrinos and quarks. Historically, the ratio of the neutral to charged current cross section has been considered as a measurement of the weak mixing angle, as it depends on $\sin^2 \theta_w$. Since the weak mixing angle is measured very precisely by other facilities such as LEP~\cite{Schael:2013ita}, however, we choose to assume no deviations from precision electroweak physics in the SM, and instead place limits on BSM interactions. As fully $SU(2) \times U(1)$-symmetric interactions typically face strong constraints from processes involving charged leptons, we focus on the usual NSI~\cite{Coloma:2019mbs}
\be
\!\!\!\mathcal{L} \supset -\sqrt{2} G_F \!\!\! \sum_{f,\alpha,\beta}^{\phantom{1}}  [\bar{\nu}_\alpha  \gamma^\mu P_L \nu_\beta] [\epsilon_{\alpha\beta}^{f,V} \bar{f} \gamma_\mu  f + \epsilon_{\alpha\beta}^{f,A} \bar{f} \gamma_\mu \gamma^5  f]\!
\ee
where $f = u, d$ and $\alpha, \beta = e, \mu, \tau$. These interactions would interfere with $Z$ exchange, affecting the neutrino neutral current cross section. Data on neutrino oscillations~\cite{Wolfenstein:1977ue} and coherent neutrino-nucleus scattering~\cite{Barranco:2005yy} probe the vector couplings $\epsilon^{f,V}_{\alpha\beta}$ efficiently but are not sensitive to their axial counterparts that only couple to net spin. By contrast, high-energy experiments can probe NSI regardless of the underlying spin structure~\cite{Altmannshofer:2018xyo, Babu:2020nna, Liu:2020emq}.

In passing, we remark that while the validity of any effective operator treatment breaks down at sufficiently high energies, the momentum transfers that we consider are of order $\sqrt{2 m_N E_\nu} \lesssim v$, where $v=246~\gev$ is the electroweak vacuum expectation value. We will obtain limits on the NSI parameters that are less than $\mathcal{O}(1)$, corresponding to operator suppression scales above the electroweak scale. At even higher energies, of course, a full UV completion of the NSI should be considered~\cite{Babu:2020nna}. It would be interesting to examine the sensitivity of FASER$\nu$ neutrino NC scattering measurements to light mediators, where we would expect different kinematics from NC scattering in the SM.

To limit NSI, we anticipate a FASER$\nu$ measurement of the neutrino neutral current cross section as in Fig.~\ref{fig:nsilimits}, in conjunction with a charged current cross section measurement~\cite{Abreu:2019yak}. We take the ratio of the neutral current to the charged current cross section assuming that the flux uncertainties will largely cancel, with the main remaining considered sources of error being the statistics on the neutral current events and the uncertainty on the background.
Following the discussion above, other uncertainties such as the neutral hadron flux and energy estimation are assumed to be subdominant in the cross section ratio. In particular, FASER will directly measure the muon flux and energy spectrum once it turns on which will reduce the neutral hadron flux uncertainties. By performing a $\chi^2$ fit using the cross section ratio in each energy bin as input, we obtain the overall expected NSI sensitivity. Throughout, we make the simplifying assumption that all of the incoming neutrinos are muon (anti)neutrinos. Weaker bounds could, in principle, be obtained on NSIs involving electron neutrinos using the subdominant $\nu_e$ flux.

Our projected sensitivity is shown in \cref{fig:nsilimits}. We also show the limits obtained from taking the ratio of the NC to CC cross-sections at CHARM~\cite{Allaby:1987vr}, as well as the current bound on the vector NSI couplings from oscillation and COHERENT data~\cite{Coloma:2019mbs}. We note that CHARM probes a different combination of the up and down quark NSIs because the limits come from neutrino scattering, whereas at FASER$\nu$ we have a combined constraint from neutrinos and antineutrinos. In summary, we find that FASER$\nu$ has the potential to provide competitive NSI sensitivity, particularly in the axial case where bounds from oscillation and coherent scattering experiments do not exist.

\section{Outlook}
\label{sec:outlook} 

While LHC neutrinos have never been directly detected, \fasernu will provide the ability to probe their interactions for the first time. Measurements of neutrino cross sections at TeV-scale neutrino energies will fill a gap between lower energy laboratory experiments and astrophysical neutrino data. Neutral current scattering is significantly more difficult to observe than charged current scattering, owing to the final state neutrino that carries away much of the energy of the interaction. At FASER$\nu$, there is a significant background from neutral hadrons induced by muons from the LHC.

We have demonstrated that the neutral hadron background to neutrino neutral current scattering in \fasernu can be significantly reduced for neutrino energies $\gtrsim 100~\mathrm{GeV}$. Furthermore, we have shown that the energy of the incoming neutrino can be estimated from the measured particles exiting the interaction vertex. The precision of our energy estimation procedure for neutral current scattering is comparable to that which could be obtained for charged current scattering. In both our handling of the background and our estimation of the neutrino energies, we have used neural networks to make maximal use of the available kinematic information in each event. We have identified areas where further work is warranted to maximize the power of a full experimental analysis, in particular, the improvement of simulation tools for neutrino DIS at TeV energies and the quantification of associated uncertainties. 

The NC cross section measurement here would serve as a test of whether neutrinos interact as predicted by the SM, and can thus be used to test new couplings between neutrinos and quarks. We have interpreted our projected cross section measurements in terms of limits on neutrino NSIs, finding sensitivities that are competitive with other experiments. In particular, we obtain limits on axial NSIs, which are not constrained in any significant manner by oscillation or coherent scattering data.

As the most weakly interacting particles in the SM, there is still much to be learned about neutrinos. We have extended the potential of the LHC to test neutrino couplings by considering NC scattering at \fasernu. Taken in the broader context of data from dedicated laboratory and astrophysical neutrino facilities, we hope that collider studies of neutral current scattering will lead to an increased understanding of the neutrino sector.

\section*{Acknowledgements}

We thank Wolfgang Altmannshofer, Akitaka Ariga, Tomoko Ariga, Kaladi Babu, Michele Tammaro, Sebastian Trojanowski, and Jure Zupan for useful discussions.  We are also grateful to the authors and maintainers of many open-source software packages, including
\texttt{CRMC}~\cite{CRMC},
\texttt{EPOS-LHC}~\cite{Pierog:2013ria},
\texttt{GENIE}~\cite{Andreopoulos:2015wxa}, 
\texttt{Jupyter} notebooks~\cite{soton403913}, 
\texttt{Keras}~\cite{keras},
\texttt{Matplotlib}~\cite{Hunter:2007}, 
\texttt{NumPy}~\cite{numpy}, 
\texttt{PyHepMC}~\cite{buckley_andy_2018_1304136},
\texttt{Pythia~8}~\cite{Sjostrand:2014zea},
\texttt{QGSJET-II-04}~\cite{Ostapchenko:2010vb},
\texttt{scikit-hep}~\cite{Rodrigues:2019nct}
\texttt{scikit-learn}~\cite{scikit-learn}, and
\texttt{uproot}~\cite{jim_pivarski_2019_3256257}.
F.K. is supported by the Department of Energy under Grant No. DE-AC02-76SF00515. The work of A.I and R.M.A is supported by the U.S.~Department of Energy under Grant No. DE-SC0016013. Some of the computing for this project was performed at the High Performance Computing Center at Oklahoma State University supported in part through the National Science Foundation Grant OIA-1301789, with thanks to Evan Linde for technical assistance.

\appendix

\section{Comparison of Different Generators}
\label{sec:appendix-gnrts}

\begin{figure*}[th!]
\includegraphics[width=1\textwidth]{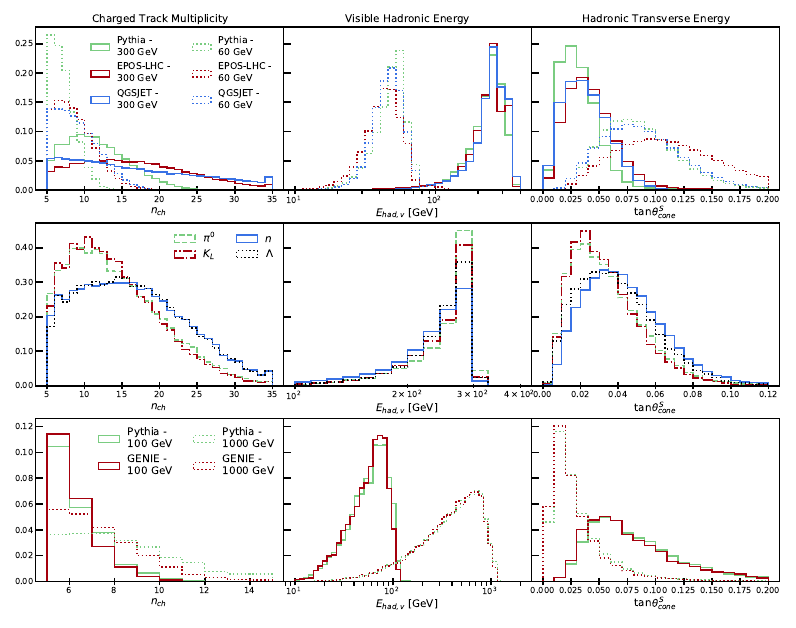}
\caption{Comparison of 3 observables for NH interactions generated using \texttt{Pythia~8}, \texttt{EPOS-LHC}, and \texttt{QGSJET-II-04} and NC interactions using \texttt{Pythia~8}, and \texttt{GENIE}. \textbf{Top:} NH interactions at $60~\gev$, $300~\gev$. \textbf{Middle:} Different NHs (mesons and baryons) generated via \texttt{EPOS-LHC} at $300~\gev$. \textbf{Bottom:} NC interactions at $100~\gev$, $1~\tev$ generated via \texttt{Pythia~8}, and \texttt{GENIE}. See text for more details.}
\label{fig:bgrnd-obs}
\end{figure*}

Throughout this work we have used signal and background samples generated using \texttt{Pythia~8}. Though reasonable agreement was found between various generators for background samples in \cref{fig:obs}, here we explore generator differences in more detail. Currently, generators do not provide uncertainties for their hadronic interaction models. Also, no generator has been tuned to neutrino interactions at the energies of interest in this work. We therefore compare between \texttt{Pythia~8}~\cite{Sjostrand:2006za, Sjostrand:2014zea} (using the \texttt{Monash} tune~\cite{Skands:2014pea} and the \texttt{nCTEQ15} nPDF for tungsten~\cite{Kovarik:2015cma, Kusina:2015vfa}) and \texttt{GENIE}~\cite{Andreopoulos:2009rq, Andreopoulos:2015wxa} for neutrino interactions and \texttt{Pythia~8}, \texttt{EPOS-LHC}~\cite{Pierog:2013ria}, and \texttt{QGSJET-II-04}~\cite{Ostapchenko:2010vb} for neutral hadron interactions to obtain a first qualitative understanding of the sizes of these uncertainties.

We first look at the differences between \texttt{Pythia~8}, \texttt{EPOS-LHC}, and \texttt{QGSJET} for neutral hadron interactions. The latter have been carefully tuned to a variety of data from collider, heavy ion, fixed target cosmic ray experiments. We show three relevant observables for neutral hadron energies of $60~\gev$ and $300~\gev$ (top row in \cref{fig:bgrnd-obs}). The differences between generators are greatest for the charged track multiplicity, where \texttt{Pythia~8} predicts more events with small numbers of charged tracks than \texttt{EPOS-LHC} or \texttt{QGSJET}. In \cref{fig:obs}, we use the range of generator predictions for the observables to draw the shaded uncertainty bands, using the physical energy spectrum for neutral hadron events. This spectrum peaks around $E \sim 60~\gev$, and events at lower energy tend to fail the neutral vertex identification criteria due to the lack of $\geq 5$ charged tracks. The shape of the charged track multiplicity distribution at very low $n_\text{ch}$ can thus affect the neutral vertex identification efficiency of \cref{fig:classification}. Measurements from dedicated beam dump experiments such as DsTau~\cite{Aoki:2019jry} and NA61~\cite{Gazdzicki:995681} could help in tuning generators in the relevant kinematic regime to reduce this background uncertainty. In particular, DsTau\cite{Aoki:2019jry} will use the high energy SPS beam and a tungsten based emulsion detector, and so, would be able to provide data in an environment that is very similar to that at FASER$\nu$.

In the second row we compare the same distributions for interactions of different neutral hadrons, $\pi^0$, $K_{\tiny{L}}$, $n$ and $\Lambda$, at $300~\gev$, as generated with \texttt{EPOS-LHC}. The event shapes for $\pi^0$ and $K_{\tiny{L}}$ (meson) are similar to each other, as are those for $n$ and $\Lambda$ (baryon) lines, but the two groups differ from each other. This shows that there is a slight difference between mesons and baryons. The difference is small enough to justify our simplified assumption of simulating all interactions of neutral hadrons as neutrons. We note, however, that these differences should be taken into account for a full experimental analysis. Also, the similarities between the $\Lambda$ and $n$ event shapes, as well as those for $K_{\tiny{L}}$ and $\pi^0$, allow us to conclude that the strange quark content is not important for our simulations.

The last row of \cref{fig:bgrnd-obs} shows NC interactions generated using \texttt{Pythia~8} and \texttt{GENIE}, at energies of $100~\gev$ and $1~\tev$. In spite of \texttt{Pythia~8} not including final state interactions, we see good agreement between the two. So far, no generator has been tuned at these energies for neutrino interactions and hence care must be taken while interpreting these results as a measure of our background uncertainty.


\bibliography{references}

\end{document}